\documentclass[prl,preprint,amsmath,amssymb,notitlepage]{revtex4-1}

\usepackage{graphicx}
\usepackage{amssymb}
\usepackage{amsfonts}
\usepackage{amsmath}
\usepackage{amsbsy}
\usepackage{natbib}
\usepackage{comment}

\usepackage{color}
\usepackage{float}

\usepackage[colorlinks=true, urlcolor=blue, citecolor=blue, linkcolor=blue]{hyperref}

\renewcommand{\vec}[1]{\mbox{\boldmath$\mathrm{#1}$}}
\let\sb=_ \catcode`\_=\active \def_#1{\ensuremath \sb{\rm#1}}

\renewcommand{\vec}[1]{\mbox{\boldmath$\mathrm{#1}$}}
\newcommand{\be}{\begin{equation}}
\newcommand{\ee}{\end{equation}}
\newcommand{\ben}{\begin{eqnarray}}
\newcommand{\een}{\end{eqnarray}}

\newcommand{\dblue}[1]{{\color{black} #1}}
\newcommand{\dbluend}[1]{{\color{black} #1}}

\begin{document}


\title{\dblue{Floquet-engineering the exceptional points in parity-time-symmetric  magnonics }}

\author{Xi-guang Wang$^{1}$, Lu-lu Zeng$^1$, Guang-hua Guo$^{1}$, Jamal Berakdar$^{2*}$}

\address{$^1$ School of Physics and Electronics, Central South University, Changsha 410083, China \\
	$^2$ Institut f\"ur Physik, Martin-Luther Universit\"at Halle-Wittenberg, 06099 Halle/Saale, Germany \\
    $^*$ email: jamal.berakdar@physik.uni-halle.de}

\date{\today}

\begin{abstract}
Magnons serve as a testing ground for fundamental aspects of Hermitian and non-Hermitian wave mechanics and are of high relevance for information technology. This study presents setups for realizing 
 spatio-temporally driven  parity-time (PT) symmetric magnonics based on coupled  magnetic waveguides and magnonic crystals.  \dblue{
A charge current in a metal layer with strong spin-orbit coupling sandwiched between  two  insulating   magnetic waveguides   leads to gain or loss in the magnon amplitude depending on the directions of the magnetization and the charge currents. When gain in one wave guide is balanced by loss in the other waveguide a PT-symmetric system hosting non-Hermitian degeneracies  (or   exceptional points (EPs)) is realized. 
 For AC current multiple  EPs  appear   for a certain gain/loss strength and  mark the boundaries between  the preserved PT-symmetry  and the broken PT-symmetry  phases. The number of islands of broken PT-symmetry  phases and their extensions is tunable by the frequency and the strength of the spacer current. 
 At EP and beyond, the induced and amplified magnetization oscillations are strong and self-sustained. In particular, these magnetization auto-oscillations in  broken PT-symmetry phase   occur at low current densities and  do not require further adjustments such as   tilt angle between electric polarization and equilibrium magnetization direction in spin-torque oscillators, 
pointing to a new design of these oscillators and their utilization in computing and sensoric. It is also shown how the  periodic gain/loss mechanism allows for  the generation of high-frequency spin waves with low-frequency currents. For spatially-periodic gain/loss acting on a magnonic crystal,   magnon modes  approaching each other at the Brillouin-zone boundaries are highly susceptible to PT-symmetry, allowing for a  wave-vector-resolved experimental realization  at very low  currents.}
\end{abstract}

\maketitle

\begin{figure}[htbp]
	\includegraphics[width=0.9\textwidth]{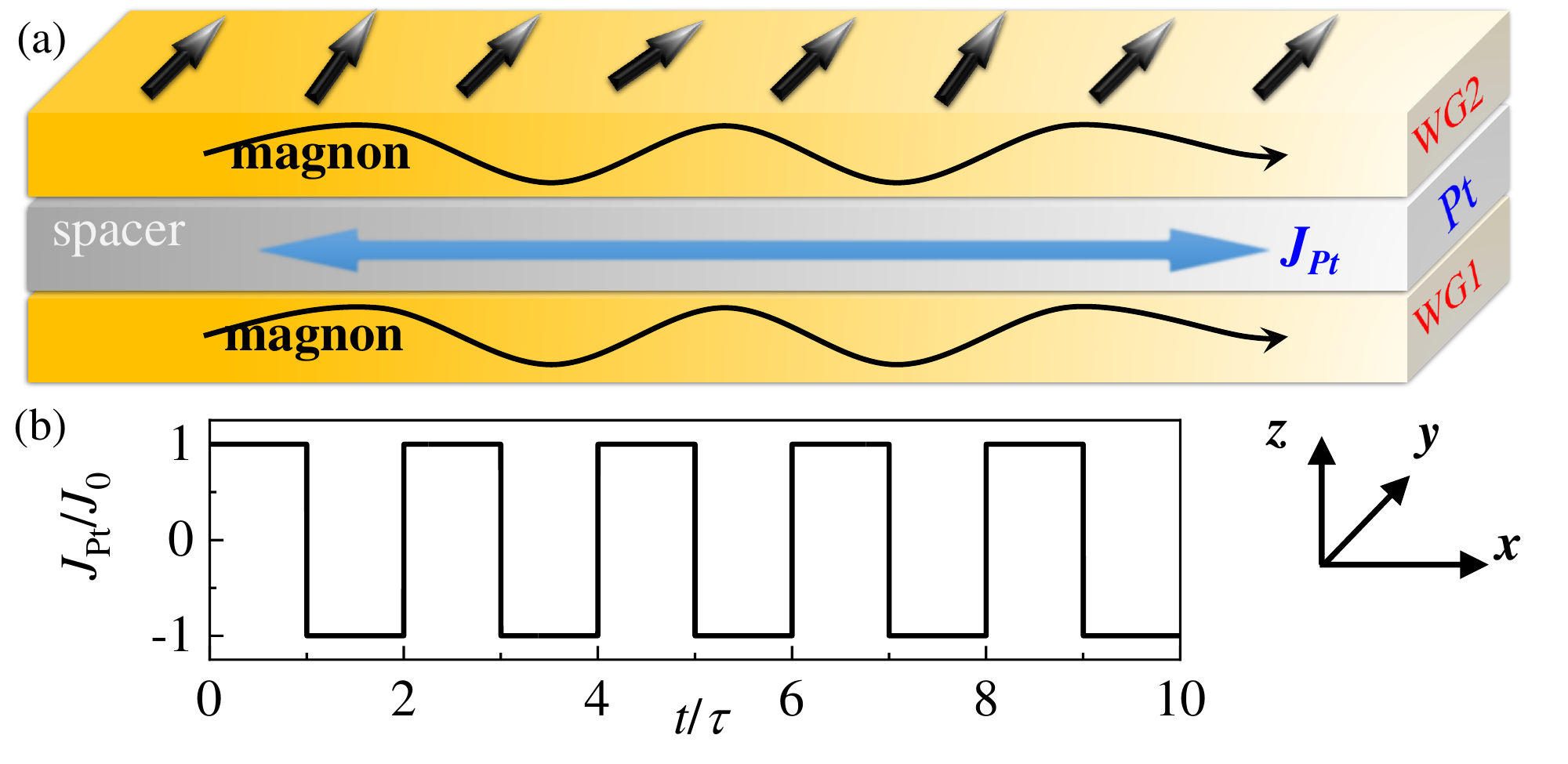}
	\caption{\label{model} 	(a) Schematic of the RKKY-coupled magnonic waveguides with time-periodic PT symmetry. Two magnetic  films serve as the magnon waveguides (WG1 and WG2) and are coupled ferromagnetically by a nanoscale conductive spacer (for instance Pt) with a spin Hall angle. Injecting into the spacer a time-varying charge current $ J_{Pt} $ results in opposite spin-orbit torques (SOTs) acting on the magnetic dynamics in WG1 and WG2 and leading to  time-varying magnonic gain/loss in WG1/WG2. (b)  The time-periodic current $ J_{Pt} $ with period $ T = 2 \tau $  used in the simulations. }
\end{figure}

\textit{Introduction}:  Long-wave length elementary excitations around a magnetically ordered stable state in an extended system are spin waves (SW) or magnons (in reference to their excitation quanta). SWs can be geometrically and magneto-statically (finite-size) quantized and steered by a variety of external probes such as magnetic and electric fields as well as by charge currents or temperature gradients.\cite{stancil2009spin,prabhakar2009spin,DEMOKRITOV2001441,Uchida2010,50007095}  Thus, SWs are well suited for use in (classical) information transfer and processing.\cite{Chumak2015,Chumak} Basic hardware elements for magnonic data channeling and processing are thereby coupled magnonic waveguides (WGs), meaning  magnetically ordered stripes that exchange power via coupling mediated by dipolar fields, or   Rudermann-Kittel-Kasuya-Yosida (RKKY) interaction.\cite{Wang2018,Wangq2020,FanYabin2020,Sadovnikov2018}\\
   Intrinsic magnetic damping may compromise the fidelity of information and cause power dissipation. A way to act on  magnetic damping externally  is offered by spin-orbit torque (SOT),
  which arises when the WG is attached to a metallic (such as Pt) layer with strong spin orbit coupling.\cite{50007095,PhysRevLett.109.096602,Garello2013,Hoffmann2013}  The magnetic damping strength $\alpha$ in the WG is then controlled by a DC bias on the metallic layer that drives a charge current density  $ \vec{J}_{Pt} $.
  The mechanism behind this effect is in short: an interfacial spin dependent scattering generates  spin accumulations  $ \vec{A} $ at the interfaces of the WG/metallic layer, the strength of which is set by the spin Hall angle. The
  spin accumulations act  with a torque (spin orbit torque (SOT)) on the magnetization density $\vec{m}$. The torque direction  is set by the charge current density and magnetization vectors.  Flipping the direction of $\vec{m}$ or $ \vec{J}_{Pt} $ changes the sign of $ \vec{T} $. As for the SWs dynamics, SOT 
  has a field-like and damping/antidamping-type effects. It is possible to realize a situation where the metallic layer is sandwiched between two coupled magnonic WGs resulting in damping (magnonic loss)  in one WG and antidamping (magnonic gain)  in the other WG.\cite{Wangxinc2020,Yu2020,PhysRevApplied034050,wangxiapl2020,PhysRevApplied.18.024080} This is a typical situation of gain/loss setup, as discussed in  connection with 
  parity-time (PT) symmetric systems
 \cite{PhysRevLett.80.5243,PhysRevLett.89.270401,Bender2007,Feng2014,Ganainy2018,Ruter2010,Regensburger2012,Miri2019,Zhu2014,Fleury2015,Assawaworrarit2017,Chen2018,Lee2015,Zhang2017,Yang2018,Galda2019,Liu2019, Wangxinc2020, PhysRevApplied034050, wangxiapl2020,Sui2022,PhysRevLett.80.5243,PhysRevLett.89.270401,Bender2007}
 which have been  investigated in optics\cite{Feng2014,Ganainy2018,Ruter2010,Regensburger2012,Miri2019}, acoustics\cite{Zhu2014,Fleury2015}, electronics\cite{Assawaworrarit2017,Chen2018} and spintronic\cite{Lee2015,Zhang2017,Yang2018,Galda2019,Liu2019, Wangxinc2020, PhysRevApplied034050, wangxiapl2020,Sui2022,Yu2020,PhysRevApplied.18.024080}. 
  Indeed, it is demonstrated mathematically \cite{Wangxinc2020}, how the dynamics in 
 coupled gain/loss magnonic WGs (the system studied here) can  be mapped to PT-symmetric non-Hermitian Hamiltonian. Such system can be driven externally to the 
 non-Hermitian degeneracy, or exceptional point (EP)  by changing SOT, meaning $ J_{Pt} $.\\
 EP can be viewed 
 as a separation point between the PT symmetry broken/or preserving phases \cite{Feng2014, Ganainy2018, Ruter2010,Miri2019,PhysRevApplied034050,Hodaei2016,Wiersig2020,Zhang2018,Lai2019,PhysRevLett.112.203901}. Approaching the EP, different eigenmodes collapse   changing significantly the mode propagations.  Above the EP, in the  PT-symmetry-broken  phase, the mode propagation can be amplified (attenuated) under gain (loss). These special features can be exploited    for designing new types of devices for information processing and for sensoric. 
 Experimentally, the existence of EP has been validated in two coupled magnets with different damping \cite{Liu2019}, and where the damping is imparted   by  laser pulses \cite{PhysRevApplied.18.024073}. \\
 So far,  gain and loss in magnon amplitude were  considered to be homogeneous in time and space across the magnonic WGs. 
In the present work, we show that   PT-symmetric magnonic WGs with time or/and space-varying gain and loss show qualitatively new features that are of direct relevance for applications.
As sketched in Fig.\ref{model}, such systems can be fabricated by nanostructuring the Pt layer or by a time-dependent  voltage which drives then a time-dependent  
$ J_{Pt} $. An appropriate framework for analyzing the dynamics in these cases is the 
 Floquet  or the Bloch-state approach applied to PT-symmetric non-Hermitian  dynamics.
Performing the analysis, we  identify multiple EPs from Floquet quasienergies where  magnon amplification is induced.\\ Comparing with a constant magnonic gain and loss,  the charge current densities needed to approach the EPs are   much smaller, and are tunable with   the time period of $J_{Pt} $.
For space periodic gain and loss,  EP's $ J_{Pt} $ becomes  very low (compared to typical current densities needed to reach EP in homogeneous PT-symmetric systems \cite{Wangxinc2020} or to induce magnetic switching/oscillation) when two magnon modes approach each other, which typically occurs when folding  modes at the   Brillouin-zone (BZ) boundaries.
In contrast to WGs with homogeneous gain/loss, where  spin reversal above EP occurs, in our case  magnon amplification around the EPs leads eventually   to   self-sustained oscillation. Recalling the well-documented case of  SOT induced magnetization auto-oscillation which depends on the tilt angle between the electric polarization and the equilibrium magnetization direction \cite{Haidar2019,Fulara2019,Houssameddine2007,Kaka2005}, our finding points to a new way for  realizing SOT oscillators without tilt-angle adjustments and with the high sensitivity to external probes  akin to systems with EPs \cite{PhysRevApplied.18.024080}.

 \textit{Model}: The considered  magnonic PT-symmetric coupled WGs with time-periodic gain/loss is illustrated in Fig. \ref{model}. Two insulating magnetic layers are  initially magnetized along $ +y $ direction and coupled,  via the RKKY interaction that acts through \dbluend{a spacer}  (Fig. \ref{model}(a)). \dblue{Due to the spin Hall effect,} a time-varying charge current in the spacer results in SOTs $ \vec{T}_{1(2)} = \gamma c_J \vec{m}_{1(2)} \times (\pm \vec{y}) \times \vec{m}_{1(2)} $.\cite{PhysRevLett.109.096602,Garello2013,Hoffmann2013} The coupled (linear and nonlinear) magnetic dynamics is governed by  Landau-Lifshitz-Gilbert (LLG) equations,
\begin{equation}
\begin{aligned}
\displaystyle 
\label{llg} \frac{\partial \vec{m}_p}{\partial t} = -\gamma \vec{m}_p \times \vec{H}_{eff,p}  + \alpha \vec{m}_p \times \frac{\partial \vec{m}_p}{\partial t} + \vec{T}_p,
\end{aligned}
\end{equation}
 $ p = 1,2 $ enumerate WG1 and WG2. $ \gamma $ is the gyromagnetic ratio, and $ \alpha $ is the intrinsic Gilbert damping. The effective field $ \vec{H}_{eff,p} = \frac{2 A_{ex}}{\mu_0 M_s} \nabla^2 \vec{m}_p + \frac{J_F}{\mu_0 M_s t_p} \vec{m}_{p'} + H_0 \vec{y} $ consists of the internal exchange field (with an exchange constant $ A_{ex} $), the RKKY interlayer coupling field (with a coupling constant $ J_F $), and the external field $ H_0 $. $ M_s $ is the saturation magnetization. $ p, p' = 1, 2 $ and $ p \ne p' $. $ t_p $ is the $ p $th layer thickness and $ \mu_0 $ is the vacuum permeability. \dblue{The SOT strength coefficient $ c_J = \frac{S \theta_{SH} \hbar J_{Pt}} { 2 \mu_0 e t_p M_s } $ is proportional to the charge current density $J_{Pt}$, the spin-Hall angle $\theta_{SH}$ of the spacer, and the WG/spacer interface transparency $ S $.}
 
 \begin{figure}[htbp]
 	\includegraphics[width=1\textwidth]{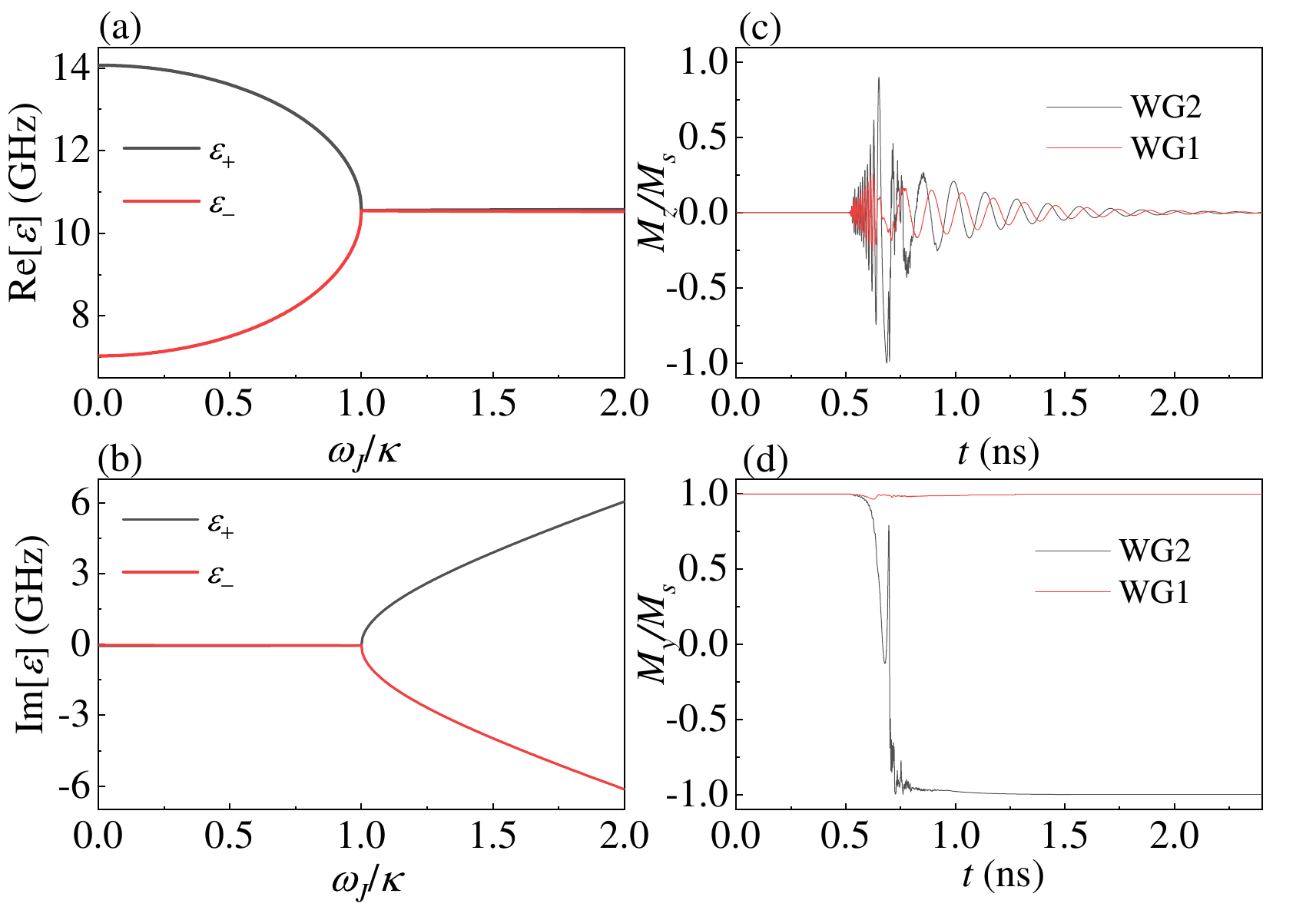}
 	\caption{\label{constant}  \dblue{(a) real and (b) imaginary parts of quasienergies $ \epsilon_{\pm} $ (Eq. (\ref{eigen1})) as functions of SOT coupling strength $ \omega_J $ at $ k_x = 0 $ and constant current in the spacer. For $ \omega_J / \kappa = 1.1$, time-dependent $ M_z$ (c) and  $M_y$ (d) at $ x = 2000 $ nm in WG1 and WG2 (obtained from the numerical simulation based on Eq. (\ref{llg})). Spin waves are excited by sinc pulse $ h(t) = h_a \vec{z} \sin(\omega_F t) / (\omega_F t) $ with amplitude $ h_a = 1 \times 10^5 $ A/m and frequency range $50$ GHz applied locally to the region $ x = 0 $.} }
 \end{figure} 
 
 In numerical calculations, we adopt the following parameters for \dblue{Yttrium–Iron–Garnet (YIG)}: saturation magnetization $ M_s = 1.4 \times 10^5 $ A/m, exchange constant $ A_{ex} = 3 \times 10^{-12} $ J/m. \dblue{The Gilbert damping  $\alpha$ can be experimentally varied   within a wide range  via material engineering \cite{PATI2020104821} ($\alpha=(6.15 \pm 1.5) \times 10^{-5}$ is reported in \cite{Hauser2016}), below we use $\alpha=0.004 $}. The interlayer exchange constant is $ J_{F} = 9 \times 10^{-5} $ J/m$ ^2 $, and the exchange field amplitude is $ \frac{J_F}{\mu_0 M_s t_p} \approxeq 1 \times 10^5 $ A/m with $ t_p = 4 $ nm. A sufficiently strong magnetic field $ H_0 = 2 \times 10^5 $ A/m is applied along the $ +y $ (or $ +x $) direction to drive the WGs to the saturated state.

We will perform full-fledged  numerical simulations and to analyze the numerical results in certain regimes, we setup  a linearized analytical model. 
For analytical modeling   we consider small deviations of $ \vec{m}_{s,p} = (\delta m_{x,p}, 0, \delta m_{z,p}) $ around the equilibrium $ \vec{m}_{0,p} = \vec{y} $. Defining $ \psi_p = \delta m_{x,p} + i \delta m_{z,p} $, we deduce the coupled SW equation under the linear assumption ($\|\vec{m}_{s,p}\|\ll 1$), 
\begin{equation}
\begin{small}
\begin{aligned}
\displaystyle 
\label{coupled} i\frac{\partial \psi_1}{\partial t} - [(\omega_0 - \alpha \omega_{J,t}(t)) - i (\omega_J + \alpha \omega_0)]\psi_1 + q \psi_2 = 0,\\
                i\frac{\partial \psi_2}{\partial t} - [(\omega_0 + \alpha \omega_{J,t}(t)) + i (\omega_J - \alpha \omega_0)]\psi_2 + q \psi_1 = 0.
\end{aligned}
\end{small}
\end{equation}
The WG dispersion of the intrinsic frequency  $ \omega_0 $ with respect to the wave vector $k_x$ reads $ \omega_0 (k_x) = \frac{\gamma}{1+\alpha^2}(H_0 + \frac{2A_{ex}}{\mu_0 M_s} k_x^2 + \frac{J_F}{\mu_0 M_s t_p}) $, the time-dependent \dblue{SOT term} related to the gain-loss mechanism is  $ \omega_{J,t}(t) = \frac{\gamma c_J(t)}{1+\alpha^2} $, and the RKKY-related (static) coupling term is $ q = (1-i\alpha) \kappa $ with $ \kappa = \frac{\gamma J_F}{(1+\alpha^2)\mu_0 M_s t_p} $. \dblue{With the Hall angle $ \theta_{SH} = 0.06 $ and interface transparency $S = 0.25$, the value of $\omega_J = \kappa$ corresponds to a charge current density of $J_e = 9 \times 10^8 {\rm A/cm^2}$}.

 \textit{Results}: When the bias voltage on the spacer is alternating with period $ T $,  we have  $ \omega_{J,t}(t + T) = \omega_{J,t}(t) $. For clarity we study a simple periodic step-function  with alternating polarity (cf. Fig.\ref{model}(b)) such that if $ \omega_{J,t}( 0 \leq t \leq \tau ) = \omega_J  $ and  $ \omega_{J,t}( \tau \leq t \leq T ) = -\omega_J $. Thus, $ T=2\tau  $.
 We consider   the case where the gain compensates for the  loss (meaning the two  WGs are of the same material). Preparing the system to be in the state   $ (\psi_{1}^0, \psi_{2}^0) $, and starting with positive (negative) SOT polarity from Eq. (\ref{coupled}), the evolved state after time $ \tau $ reads 
\begin{equation}
\begin{aligned}
\displaystyle 
\label{mat} \left( \begin{matrix}
\psi_1 \\
\psi_2
\end{matrix}\right)
= \hat{M}_{P(N)}(\tau)
\left( \begin{matrix}
\psi_1^0 \\
\psi_2^0
\end{matrix}\right).
\end{aligned}
\end{equation}
For positive (negative) $ \omega_J $ we have
\begin{widetext}
	\begin{equation}
\begin{small}
\begin{aligned}
\displaystyle 
\label{mat1} M_{P} = \frac{e^{-i w_e}}{2 d}
\left( \begin{matrix}
e^{-i d_e}(d-i\omega_J) + e^{i d_e}(d+i\omega_J) && (e^{i d_e}-e^{-i d_e})\kappa \\
(e^{i d_e}-e^{-i d_e})\kappa && e^{-i d_e}(d+i\omega_J) + e^{i d_e}(d-i\omega_J)
\end{matrix}\right),\\
M_{N} = \frac{e^{-i w_e}}{2 d}
\left( \begin{matrix}
e^{-i d_e}(d+i\omega_J) + e^{i d_e}(d-i\omega_J) && (e^{i d_e}-e^{-i d_e})\kappa \\
(e^{i d_e}-e^{-i d_e})\kappa && e^{-i d_e}(d-i\omega_J) + e^{i d_e}(d+i\omega_J)
\end{matrix}\right).
\end{aligned}
\end{small}
\end{equation}
\end{widetext}
\dblue{With the spacer current angular frequency $\omega_F=\pi/\tau$, we introduced $ d = \sqrt{\kappa^2 - \omega_J^2} $, $ w_e = \pi (1 - i\alpha)\omega_0/\omega_F $, $ d_e = \pi (1 - i\alpha)   d/\omega_F $}. The combined propagation matrix after one period is $ \hat{M}(T) = \hat{M}_P(\tau) \hat{M}_N(\tau) $. For periodic driving,  Floquet's  theorem states for    solutions      $ e^{-i \epsilon_{\pm} t} \phi_{\pm} (t)$  that the function (Floquet state) $\phi_{\pm}(t)=\phi_{\pm}(t+T)$, and the Floquet's quasienergy $ \epsilon_{\pm}\in [0,\omega_F]  $ defined  up to multiples of $\omega_F$.

\begin{figure}[htbp]
	\includegraphics[width=1\textwidth]{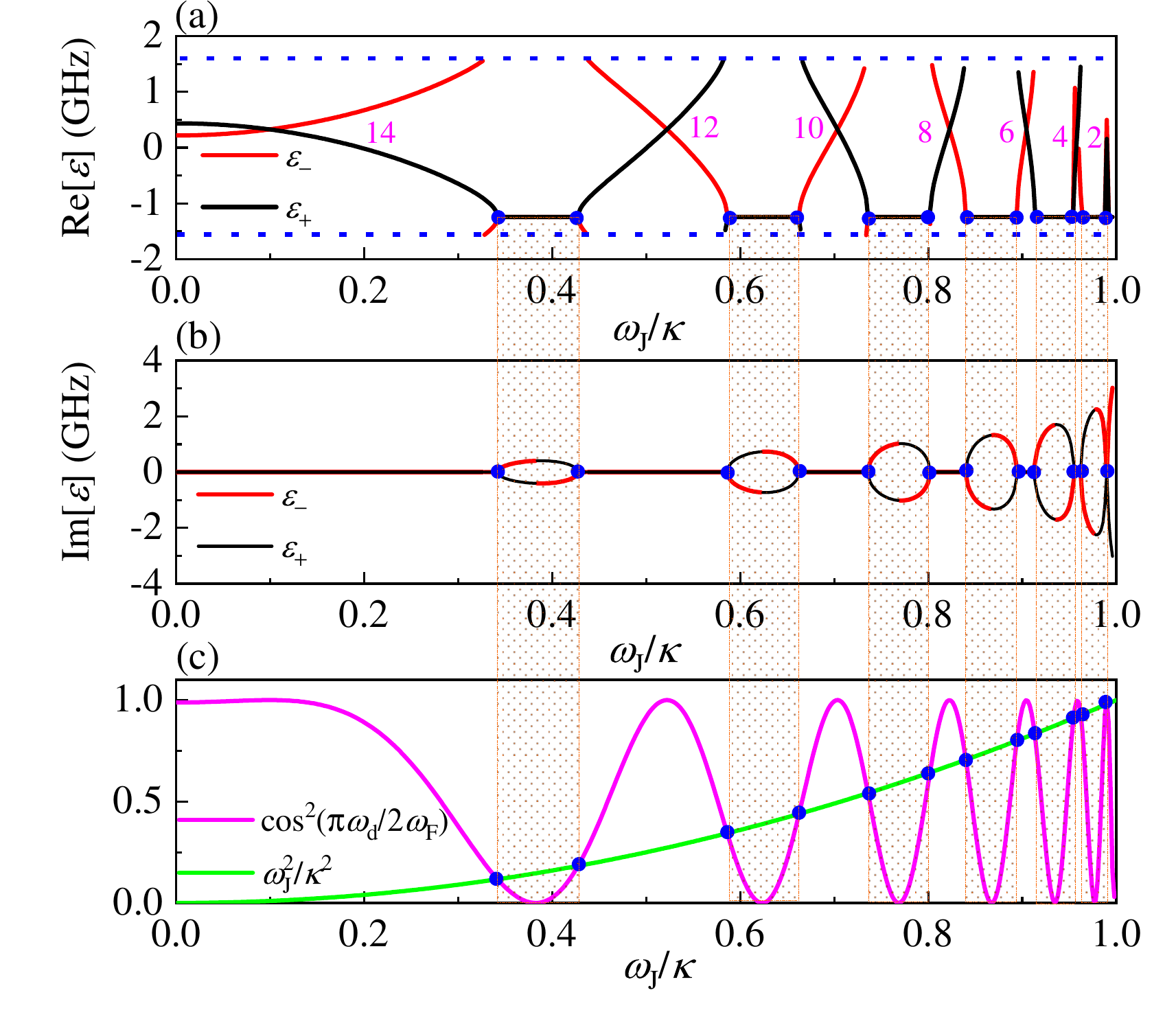}
	\caption{\label{alpha0} \dblue{ 
			 (a) real and (b) imaginary parts of Floquet quasienergies $ \epsilon_{\pm} $ (Eq. (\ref{eigen2a})) as functions of the   amplitude   $ \omega_J $ of periodic SOT with angular frequency  $ \omega_F = \pi $ GHz and $ k_x = 0 $ for $ \alpha\to 0 $. Real parts of $ \epsilon_{\pm} $ in the zone $ -\omega_F/2 $ and $ \omega_F/2 $ (marked by blue dashed lines) with $T = 2\pi/\omega_F$ being the AC charge current period. The integers (14, 12,...) in (a) are the ratio $ \frac{\omega_d}{\omega_F} $ when $ \epsilon_{\pm} $ cross. (c)  $ \frac{\omega_J^2} {\kappa^2}$ and $ \cos^2(\frac {\pi} {2} \, \frac{\omega_d}{\omega_F}) $ in Eq. (\ref{eigen2a}). Blue full dots mark Floquet EPs  conditions. Shadowed areas mark the range of  broken PT-symmetry phases. Note, the extension and the number of the broken PT-symmetry islands are controllable by the charge current density ($\propto \omega_J$) and  frequency ($\omega_F$). }}
\end{figure}

At first, we analyze the case with constant SOT, i.e., $ \tau \to \infty$. Depending on the sign of $ \omega_J $, the   evolution follows $ \hat{M}_P(t) $ or $ \hat{M}_N(t) $, and from the eigenvalues $ V_{\pm} $ of the  matrices we can obtain the complex Floquet quasienergies as 
\begin{equation}
\begin{aligned}
\displaystyle 
\label{eigen1} \epsilon_{\pm} = (1 - i\alpha)(\omega_0 \pm d).
\end{aligned}
\end{equation}
For the constant SOT, these quasienergies are exactly the same as the eigen-frequencies (optic and acoustic modes) of the coupled waveguides \dblue{with level spacing $\omega_d:=2d(1-i\alpha)$}.\cite{Wangxinc2020} \dblue{ If  $  \omega_J  \to \kappa $, then $d\to 0$ and the two modes coincide  (Fig. \ref{constant}(a-b))  signaling the occurrence of  a non-Hermitian degeneracy point   (EP) that separates} the   PT-symmetry preserved  and   broken   phases. Besides, the system becomes unstable when the imaginary part of $ \epsilon_{+}$ becomes positive \dblue{(for $\omega_J > \kappa$)}, i.e., $ 0 \leq |d| - \alpha \omega_0 $. Under a very small damping, the critical value for driving the instability is set by   EP.  A non-vanishing   $ \alpha $ shifts  the instability slightly above the EP. Here, the unstable and stable regions are separated by the line $ \omega_J = \kappa $  (for the small $ \alpha = 0.004 $).  Above the critical value (say at $ \omega_J = 1.1\kappa $), we simulate the time-dependent magnetization of WG1 and WG2. As shown in Fig. \ref{constant}(c-d), the magnetization oscillation is quickly  amplified at the beginning. The amplified oscillation renders the equilibrium magnetization of WG2 switched to the $ -y$ direction. Then, the oscillation is soon damped. \dblue{ Enhanced oscillation amplitude invalidates the linear assumption necessitating  full numerical treatment, see  Supplementary Information (SM). \cite{supp}}\\

\dblue{For the time-periodic   SOT,  
the  two complex quasienergies are deduced  as }
\begin{equation}
\begin{aligned}
\displaystyle 
\label{eigen2} &\epsilon_{\pm} = -\ln\left\{\frac{e^{-2i w_e}}{2d^2} \left[(e^{-2i d_e} + e^{2i d_e})\kappa^2 - 2\omega_J^2 \right.\right. \\
&\left. \left. \pm 2\kappa \sinh(i d_e)[2(\kappa^2 + \kappa^2 \cosh(2id_e)-2\omega_J^2)]^{1/2}\right]\right\}\omega_F/(2\pi i).
\end{aligned}
\end{equation}
\dblue{For small damping ($ \alpha\to 0 $),  $\epsilon_{\pm}$   simplify to
\begin{equation}
	\begin{aligned}
	\displaystyle 
	\label{eigen2a} 
	\epsilon_{\pm} = -\ln \left\{\frac{ e^{-2i w_e} \kappa^2}{d^2} [ 
	\cos(\frac {\pi\omega_d}{\omega_F}) -\frac{\omega_J^2}{\kappa^2} 
	%
	\pm 2 i \sin( \frac{\pi\omega_d}{2\omega_F})\,  [\cos^2 (\frac {\pi\omega_d}{2\omega_F})-\frac{\omega_J^2}{\kappa^2}]^{1/2}] \right\} \frac{\omega_F }{2\pi i}.
	\end{aligned}
\end{equation}
Following current-tunable cases are identified: \\
a) When  the frequency $\omega_F$ and amplitude $\omega_J$ of  the driving current are such that  $\cos^2 (\frac\pi 2\frac{ \omega_d} {\omega_F}) > \frac{\omega_J^2}{\kappa^2}$, the   quasienergies $\epsilon_{\pm}$ are  real and different, indicating a PT-symmetry preserved phase.\\
b) Even in this phase,  Floquet states may become  degenerate when  the level spacing  is a multiple of the driving frequency, i.e., $\omega_d=2n\omega_F$ and $n$ is an integer which resembles the standard
multiphoton resonance for weak driving \cite{LARSEN1976254} and is depicted on Fig.\ref{alpha0}(a).\\
c) At  $\cos^2 (\frac\pi 2\frac { \omega_d}{\omega_F})=\frac{\omega_J^2}{\kappa^2}$ the  modes 
  coalesce signaling  EPs in Floquet modes (called henceforth FEPs). 
  d) Above the FEP ($\cos^2 (\frac\pi 2\frac{ \omega_d} {\omega_F}) < \frac{\omega_J^2}{\kappa^2}$),   $\epsilon_{\pm}$ turn complex with the two real parts of $\epsilon_{\pm}$ being degenerate, and the two imaginary parts are different which is indicative of the broken PT-symmetry  phase.
 Since $\frac{\omega_J^2}{\kappa^2}$ is monotonous but $\cos^2 (\frac\pi 2\frac{ \omega_d} {\omega_F})$ periodic in  $\omega_J$, as we vary the current spacer strength (varying thus $\omega_J$), several broken PT-symmetry  islands appear (shaded areas in Fig.\ref{alpha0}).  FEPs  mark the boundaries of these islands.

    Numerical calculations (Fig. \ref{alpha0}) based on Eq. (\ref{eigen2a}) prove the cases a)-c). Importantly, the values of FEPs occur at much smaller $\omega_J$  (which is proportional to the current density) as the conventional   EP (at $ \omega_J = \kappa $), with the first FEP arising at $ \omega_J = 0.34\kappa $. 
     For small damping $\alpha = 2 \times 10^{-5}$, $\alpha$-induced differences in $\epsilon_{\pm}$ are negligible, (see  calculations in  SM \cite{supp}). In the PT-symmetry preserved case, the  quasienergies $\epsilon_{\pm}$ satisfy as usual $\epsilon_{\pm} = \epsilon_{\pm} \pm n \omega_F  $, with $ n = 0, \pm1,... $ (see Fig. \ref{alpha0}).}
\dblue{For larger damping, e.g. $ \alpha = 0.004$, in the broken PT-symmetry  phase  small gaps appear between the real parts, and two $\rm{Re}[\epsilon_{\pm}]$ coalesce only at the center, and two imaginary parts $\rm{Im}[\epsilon_{\pm}]$ are more separated there. Besides, the damping brings in a finite  $\rm{Im}[\epsilon_{\pm}]$ outside the broken PT-symmetry  phases. These larger finite damping induced features make the FEPs not as clear as for small damping,  still we can identify the FEPs regions and broken PT-symmetry  phase from the region with more separated $\rm{Im}[\epsilon_{\pm}]$.}

\begin{figure}[htbp]
	\includegraphics[width=1\textwidth]{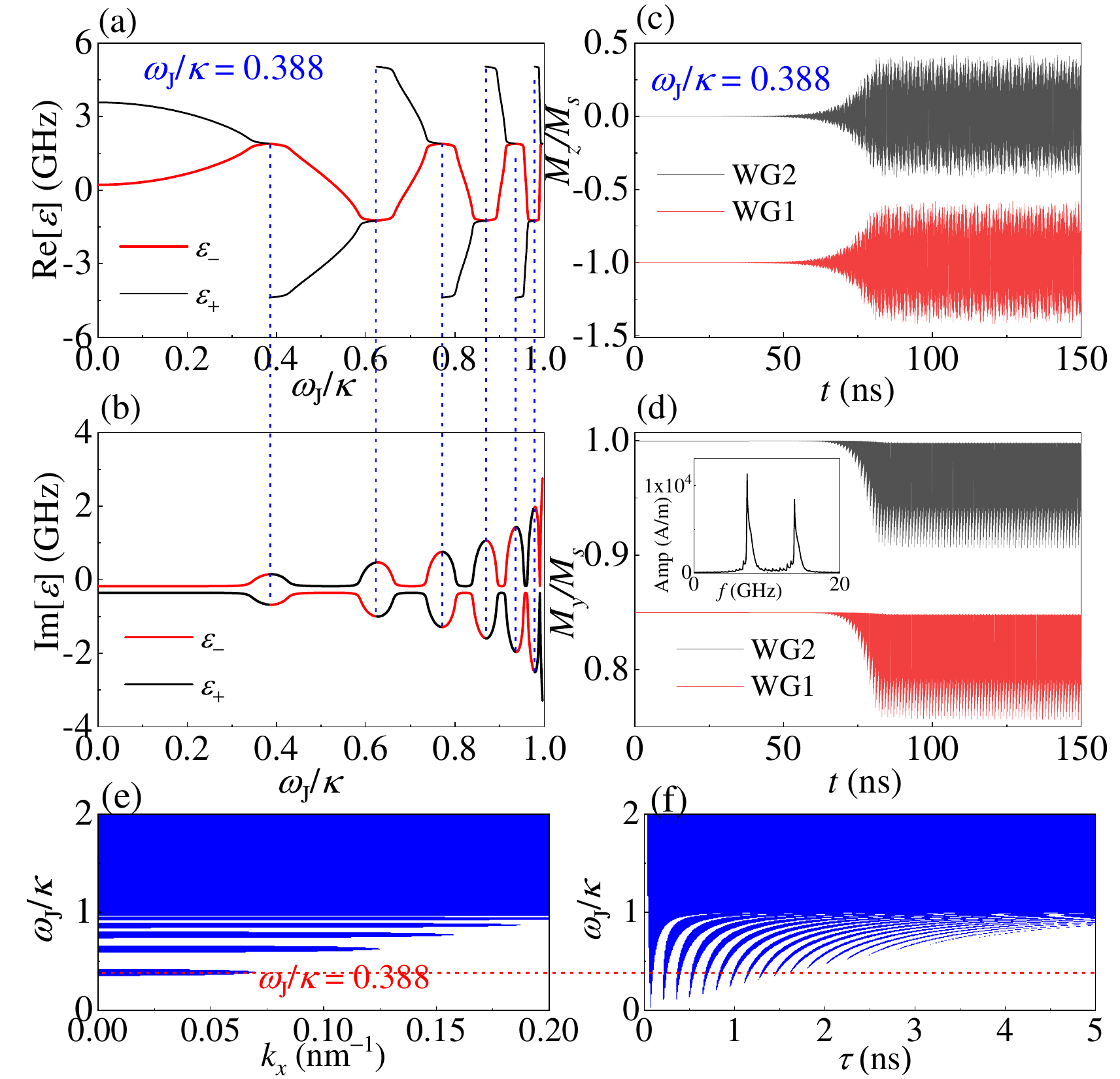}
	\caption{\label{period} For the periodic \dblue{SOT term} $ \omega_J $ (with amplitude $ \omega_J $) and $ \alpha = 0.004 $, (a) real and (b) imaginary parts of quasienergies $ \epsilon_{\pm} $ (Eq. (\ref{eigen2})) as functions of the amplitude $ \omega_J $ with the period parameter $\tau = 1$ ns and $ k_x = 0$. When $ \omega_J / \kappa = 0.388$ (inside the instability range), time-dependent (c) $ M_z (x = 2000 {\rm nm})$ and (d) averaged $M_y$ in WG1 and WG2 (obtained from the numerical simulation based on Eq. (\ref{llg})).  A sinc pulse is applied, meaning $ h(t) = h_a \vec{z} \sin(\omega_F t) / (\omega_F t) $ with the amplitude $ h_a = 1 \times 10^5 $ A/m and frequency range of $ 50$ GHz, acting locally on the region $ x = 0 $. The inset in (d) is the frequency spectrum of the magnetization oscillation in last 20 ns. \dbluend{The red curves (WG1) are in the same range with WG2 curves, which are intentionally offset for clarity.} (e-f) The stability phase diagram on the (e) $ k_x - \omega_J $ space with $ \tau = 1 $ ns, and (f) $ \tau - \omega_J $ space with $ k_x = 0$. The shaded region corresponds to the unstable oscillation.}
\end{figure}

Varying  charge current strength  (meaning gain/loss strength) the stability behavior can be controlled. Around FEPs (below $ \omega_J = \kappa $,) the imaginary parts turn positive, indicating driven instability, i.e., large-amplitude magnetization dynamics. 
 E.g., for $ \omega_J / \kappa = 0.388 $ in the instability range, the enhanced magnetization oscillation is shown in Fig. \ref{period}(c-d). 
 In contrast to  a time-constant current ($\omega_F\to 0$), the enhanced magnetization oscillation persists  long  after reaching the maximal amplitude.
 For the range with stable oscillation  (say $ 80 {\rm ns} < t < 100 {\rm ns} $), we show the frequency spectrum (see the inset of Fig. \ref{period}(d)). Mainly two resonance frequencies are excited, corresponding to the $k_x = 0$ mode of the in-phase acoustic oscillation and the out-of-phase optic oscillation in the coupled WG1 and WG2. \dblue{Importantly, the excited resonance frequencies  differ from the frequency of the spacer current (0.5 GHz) offering so a new method to generate high frequency magnons by an electric current with a very low frequency.}  Furthermore, we calculate the $ k_x - \omega_J $ (at $ \tau = 1 {\rm ns} $) and $ \tau - \omega_J $ (at $ k_x = 0$) stability diagram (Fig. \ref{period}(e-f)). For low $ k_x $, several instability islands are substantial in size; all areas shrink with  increasing  $k_x$.   For smaller current period $ T $ the separated instability areas can be enlarged. \dblue{ Important for a possible experimental realization, to drive the auto-oscillation inside the instability islands with broken PT-symmetry phase,  the required precision in $T$ is lower for the larger SOT amplitude $ \omega_J $. For example, in our estimation, to drive the instability, the smallest SOT amplitude is $\omega_J / \kappa = 0.03$, which requires the time period $T = (14.2  \pm 0.2) \times 10^{-11} $ s, i.e., the precision is around 2 ps. $\omega_J / \kappa = 0.1$ allows for $T = (14.2 \pm 1) \times 10^{-11} $ s, and larger $\omega_J / \kappa = 0.2$ has  $T = (14.6 \pm 2) \times 10^{-11} $ s, indicating lower and lower precession requirement. } 

For comparison, we also apply  constant and periodic SOT to a single waveguide. We consider the SOT $ \vec{T} = \gamma c_J \vec{m} \times (- \vec{y}) \times \vec{m} $ and the equilibrium stable magnetization $ \vec{m}_0 = \vec{y}$. Following the same procedure discussed above,  the quasienergy for constant $\omega_J$ in a single WG is $(1 - i\alpha) (\omega_0 + i \omega_J)$. When the antidamping SOT counteracts the damping torque, the imaginary value $ \omega_J - \alpha \omega_0 > 0 $, and the equilibrium magnetization $ \vec{m}_0 = \vec{y}$ loses its stability. The conclusion agrees well with SOT-induced magnetization oscillation. But the induced oscillation is not sustained.  The magnetization is soon reversed to $ -y $ and the oscillation is damped. This is in line with  studies showing  that a sustained auto-oscillation requires a finite tilt angle between the electric polarization and the equilibrium magnetization directions.\cite{Haidar2019,Fulara2019,Houssameddine2007,Kaka2005}  For periodic $\omega_J$ in a single WG, the quasienergy becomes $ (1 - i\alpha) \omega_0 $, which is independent of the amplitude of $\omega_J$. \dblue{ The reason behind this is that the opposite SOT effects from two half periods are completely neutralized.} \dblue{Besides, the above periodic varying SOT is realized in the time domain. In the SM \cite{supp}, we also analyze a spatially-periodic gain and loss driven by SOT. As the spin-wave also experiences periodically varying gain and loss during the propagation, we identify lower EPs and auto-oscillation in the broken PT-symmetry  phase above EP. Our investigation of a system with coupled periodic loss/more-loss mechanism (SM \cite{supp}) also uncovers similar features of EP, demonstrating the versatility of magnonics when combined with PT-symmetry.}

\textit{Conclusions}: We proposed, simulated and analyzed the magnonic dynamics in  time-dependent or space-dependent periodic PT-symmetric environment. We identify  typical PT-symmetry phenomena. Via the period and the amplitude of driving fields, EPs and instability threshold can be  controlled. Different from the case with a homogeneous  gain/loss, in the instability region the induced magnetization oscillation is strong and  self-sustained, which constitutes a new effective way for realizing magnetization auto-oscillation without the need for a tilt angle in the  electric polarization with respect to the equilibrium magnetization direction in a single waveguide. This finding can be used  for designing  SOT oscillators.  For the spatially-periodic case,  EP can be reached at small SOTs or charge currents when two modes approach each other which can be achieved via an appropriate modulation of the magnonic crystal. { The  electric and magnetic non-linear response at EP points to a new type of electrically manipulable Floquet magnonic metamaterials.}
 
\textit{Acknowledgements}: This work was supported by the DFG through SFB TRR227, and Project Nr. 465098690, the National Natural Science
 Foundation of China (Grants No. 12174452, No. 12074437, No. 11704415, and No. 11674400), and the Natural Science Foundation of Hunan Province of China (Grants No. 2022JJ20050, No. 2021JJ30784, and No. 2020JJ4104), and the Central South University Innovation-Driven Research Programme (Grant No. 2023CXQD036).

\bibliographystyle{apsrev4-1}
\bibliography{PT-period}

\end{document}